\documentclass[10pt,aps,prl,twocolumn,english,superscriptaddress,aps,reprint]{revtex4}

\usepackage[LGR,T1]{fontenc}
\usepackage[utf8]{inputenc}
\usepackage{color}
\usepackage{babel}
\usepackage{textcomp}
\usepackage{multirow}
\usepackage{graphicx}
\usepackage{amsmath}
\usepackage{mathtools}
\usepackage[unicode=true,pdfusetitle,
bookmarks=true,bookmarksnumbered=true,bookmarksopen=false,
breaklinks=true,pdfborder={0 0 0},pdfborderstyle={},backref=false,colorlinks=true]
{hyperref}
\hypersetup{
	linkcolor=red, citecolor=blue,  urlcolor=blue}

\makeatletter

\ProvideTextCommand{\~}{LGR}[1]{\char126#1}

\@ifundefined{textcolor}{}
{%
	\definecolor{BLACK}{gray}{0}
	\definecolor{WHITE}{gray}{1}
	\definecolor{RED}{rgb}{1,0,0}
	\definecolor{GREEN}{rgb}{0,1,0}
	\definecolor{BLUE}{rgb}{0,0,1}
	\definecolor{CYAN}{cmyk}{1,0,0,0}
	\definecolor{MAGENTA}{cmyk}{0,1,0,0}
	\definecolor{YELLOW}{cmyk}{0,0,1,0}
}

\makeatother

\makeatother

\begin{document}

\title{ Electronic structure and magnetic properties of 3\textit{d}-4\textit{f} double perovskite material}

\author{S. Kundu}
\affiliation{Department of Physics, Indian Institute of Technology Madras, Chennai 600036, India}
\author{A. Pal}
\affiliation{Department of Physics, Indian Institute of Technology Madras, Chennai 600036, India} 
\author{Amit Chauhan} 
\affiliation{Condensed Matter Theory and Computational Lab, Department of Physics, Indian Institute of Technology Madras, Chennai 600036, India}
\affiliation{Center for Atomistic Modelling and Materials Design, Indian Institute of Technology Madras, Chennai 600036, India}
\affiliation{Functional Oxide Research Group, Indian Institute of Technology Madras, Chennai 600036, India}
\author{K. Patro}
\affiliation{Department of Physics, Indian Institute of Technology Madras, Chennai 600036, India}
\author{K. Anand}
\affiliation{Department of Physics, Indian Institute of Technology (BHU) Varanasi 221005, India}
\author{S. Rana}
\affiliation{UGC-DAE Consortium for Scientific Research, University Campus, Khandwa Road, Indore 452001}
\author{V. G. Sathe}
\affiliation{UGC-DAE Consortium for Scientific Research, University Campus, Khandwa Road, Indore 452001}
\author{Amish G. Joshi}
\affiliation{CSIR-Central Glass \& Ceramic Research Institute, Naroda Centre, 168-169 Naroda Industrial Estate, Ahmedabad 382330, India}
\author{P. Pal}
\affiliation{CSIR-Central Glass \& Ceramic Research Institute, 196, Raja S.C. Mullick Road, Kolkata – 700 032, India}
\author{K. Sethupathi}
\affiliation{Department of Physics, Indian Institute of Technology Madras, Chennai 600036, India}
\affiliation{Quantum Centre for Diamond and Emergent Materials, Indian Institute of Technology Madras, Chennai 600036, India}
\author{B. R. K. Nanda}
\email{nandab@iitm.ac.in} 
\affiliation{Condensed Matter Theory and Computational Lab, Department of Physics, Indian Institute of Technology Madras, Chennai 600036, India}
\affiliation{Center for Atomistic Modelling and Materials Design, Indian Institute of Technology Madras, Chennai 600036, India}
\affiliation{Functional Oxide Research Group, Indian Institute of Technology Madras, Chennai 600036, India}
\author{P. Khuntia}
\email{pkhuntia@iitm.ac.in}
\affiliation{Department of Physics, Indian Institute of Technology Madras, Chennai 600036, India}
\affiliation{Quantum Centre for Diamond and Emergent Materials, Indian Institute of Technology Madras, Chennai 600036, India}
\affiliation{Functional Oxide Research Group, Indian Institute of Technology Madras, Chennai 600036, India}


\begin{abstract}
	
Double perovskite based magnets wherein frustration and competition between emergent degrees of freedom are at play can lead to novel electronic and magnetic phenomena. Herein, we report the electronic structure and magnetic properties
of  an ordered double perovskite material Ho$_{2}$CoMnO$_{6}$. In the double perovskite with general class A$_{2}$BB$^{'}$O$_{6}$, the octahedral B and B$^{'}$-site has a distinct crystallographic site. The Rietveld refinement of XRD data reveal that Ho$_{2}$CoMnO$_{6}$ crystallizes in the monoclinic $P$2$_{1}/n$ space group.  The X-ray photoelectron spectroscopy confirms the charge state of cations present in this material. The temperature dependence of magnetization and specific heat exhibit a long-range ferromagnetic ordering at $T_{\rm C} \sim$ 76 K owing to the presence of super exchange interaction between Co$^{2+}$ and Mn$^{4+}$ moments. Furthermore, the magnetization isotherm at 5 K shows a  hysteresis curve that confirms ferromagnetic behavior of this double perovskite. We observed a re-entrant glassy state in the intermediate temperature regime, which is attributed to the presence of inherent anti-site disorder and competing interactions. A large magnetocaloric effect has been observed much below the ferromagnetic transition temperature. The temperature dependent Raman spectroscopy studies support the presence of spin-phonon coupling  and short range order above $T_{\rm C}$ in this double perovskite. The stabilization of magnetic ordering and charge states is further analyzed through electronic structure calculations. The latter also infers the compound to be a narrow band gap insulator with the gap arising between the lower and upper Hubbard Co-$d$ subbands. Our results demonstrate that anti-site disorder and complex 3$d$-4$f$  exchange interactions in the spin-lattice account for the observed electronic and magnetic properties in this promising double perovskite material.
\end{abstract}

\date{\today}

\maketitle

\section{Introduction}
Spin correlations and the interplay  between competing degrees of freedom can lead to exotic physical phenomena in novel magnetic  materials \cite{Balents2010,Hwang2012,Khuntia2019, Khuntia2020, Koteswararao2017, Khuntia2016}. The transition metal and rare-earth based magnetic materials are promising  candidates  to harbor exciting electronic and magnetic properties that could serve as a test bed for establishing  theoretical conjectures. In addition, it offers a viable ground  to realize the rich potentials of  novel magnets  with the aim to tailor the surprising properties of these materials to address the pressing technological challenges in material science in energy harvesting, spintronics, quantum computing, and high density data storage devices \cite{Fisher2004, Baranov2009, Pathak2014, Shivaram2014, Gruner2015, Diop2018, Arh2022, Kundu2020, Khatua2022, Kumar2015,Khuntia2009, Khuntia2013, Brando2016, Sarkar2012}. In this regard, the rare-earth double perovskite oxides \cite{Bhalla2000}, A$_{2}$BB$^{'}$O$_{6}$ (A = Rare earth ions or alkaline ions; B/B$^{'}$= transition metal ions) \cite{Blasco2016, Vasiliev2008} with rock salt type ordered structure offer a promising venue in view of their multi-functional properties owing to the  co-existence of magneto-electric coupling, magneto-dielectric effects, tunable pyroelectric, ferroelectric properties,  multiferroicity, spin polarized conductivity, and superconductivity \cite{Cheong2007,Cava1988,Kobayashi1998}. Also, double perovskite  offers a highly  flexible structure wherein  external perturbations such as chemical pressure, electric and magnetic fields can tune the underlying magnetism and spin dynamics. Furthermore, double perovskites wherein   spin correlations, frustration  and spin-orbit coupling are at play, offer a rich platform to realize exotic quantum states such as spin liquid, unconventional magnetic ordering, spin-orbit driven phenomena, and topological states with emergent excitations \cite{Cook2014,Cook2015,Dey2013,Singh2020,Khatua2021,Wiebe2003,Vasala2015}. Double perovskites are excellent candidates for potential applications such as, magnetic memory, sensors and solar cell to name a few. For example, the multiferroic material HoMnO$_{3}$ with a ferroelectric transition and large polarization of 56 mC/m$^{2}$ shows large magnetocaloric effect \cite{Hur2009, Moon2018}. 
The double perovskites R$_{2}$CoMnO$_{6}$ (R = rare earth) offer a versatile platform for hosting interesting electronic and magnetic properties, which depend on the size of R ions \cite{Truong2011} for example, the spin-phonon coupling decreases upon replacing La with smaller rare-earth ions. Recent  theoretical prediction of multiferroicity in
 Y$_{2}$NiMnO$_{6}$ \cite{Kumar2010} also indicates the significance of the type of rare-earth ion in realizing myriads of complex magnetic ordering phenomena in these double perovskites. The double perovskite materials generally crystallize into two structures either ordered or disordered based upon the B/B$^{'}$-site ordering. The ordered structure prefers to be in monoclinic space group with  $P$2$_{1}$/$n$ symmetry and the disordered one crystallizes in orthorhombic space group with $Pnma$ symmetry. The ordered double perovskites A$_{2}$BB$^{'}$O$_{6}$,  generally host ferromagnetic (FM) interaction due to superexchange  interactions. In contrast, the co-existence of B$^{3+}$/B$^{'}$$^{3+}$ ions as a disorder,  introduces competing antiferromagnetic (AFM) interactions in the spin-lattice \cite{Choudhury2012, Dass2003}. The magnetic properties of  double perovskites are well interpreted following the cationic ordering and Goodenough-Kanamori (GK) rule \cite{Booth2009, Goodenough1955a}, which suggests a ferromagnetic ground state in these systems. Whereas the  deviations from the GK rule due to cationic disorder leads to the departure from the ferromagnetism also. The double perovskites with heavier rare-earth elements show unconventional properties such as magnetization reversal and inverse exchange bias  in Er$_{2}$CoMnO$_{6}$ \cite{Banerjee2018}, negative magneto-capacitance in Yb$_{2}$CoMnO$_{6}$ \cite{Choi2017}, anisotropy magnetic properties and giant magnetocaloric effect in Tb$_{2}$CoMnO$_{6}$ \cite{Moon2018}. Y$_{2}$CoMnO$_{6}$ shows multiferroic behavior owing to the presence of competing interactions \cite{Wang2019a}  while  the steps in magnetic hysteresis loop is ascribed to anti-site disorder in this double perovskite \cite{Sharma2013, Nair2014}. In a similar vein,  La$_{2}$CoMnO$_{6}$ hosts  FM  ordering and cluster glass phenomena driven by anti-site disorder \cite{Madhogaria2019}.
The unconventional ground states of such double perovskites are consequence of the B-site cationic ordering and their nominal valence states. Most of the Co/Ni/Mn-based ordered double perovskites exhibit ferromagnetic (FM) insulating ground state, which is best modeled by the superexchange interactions among B$^{2+}$ /B$^{'4+}$ ions via adjacent oxygen ions. The complex interplay between 3\textit{d}-4\textit{f} interactions in double perovskites of type R$_{2}$CoMnO$_{6}$ can lead to rich and diverse physical phenomena  such as spin-glass, multiferroicity, Griffiths phase, magnetoresistance, exchange bias, and magnetocaloric effect. In these materials,  perturbations such as the choice of cations play a crucial role in modifying the underlying electronic and magnetic properties \cite{Vasala2015}. Double perovskites without disorder are ideal  to test theoretical conjectures, however, the experimental realization of such a structure poses a great challenge owing to anti-site disorder  and defects  in real materials. This invokes to look for new double perovskites with suitable combination of rare-earth and transition metal ions wherein the disorder could be controlled precisely, offer a viable ground to realize exotic electronic and magnetic phenomena.  In this context, the recently synthesized promising double perovskite Ho$_{2}$CoMnO$_{6}$ offers an exciting platform to explore complex magnetic ordering and electronic properties.  In this material, the Ho$^{3+}$ ion  is predicted to impact the FM ordering greatly compared to other double perovskites in view of the smaller atomic radius of Ho$^{3+}$ ion. In Ho$_{2}$CoMnO$_{6}$, competition between Co$^{2+}$-O-Mn$^{4+}$  FM  superexchange interaction and Co$^{2+}$-O-Co$^{2+}$ or  Mn$^{4+}$-O-Mn$^{4+}$  AFM interaction,  which are further modulated by the active Ho$^{3+}$ spins, could lead to interesting ground state properties. A complete picture concerning the exact charge state, electronic structure, exchange interactions, disorder and spin-phonon coupling is essential to understand the underlying mechanism that drive novel magnetism in this class of double perovskite material.

 In this work, we have investigated the structural, magnetic and electronic properties of the polycrystalline samples of a partially site-ordered double perovskite Ho$_{2}$CoMnO$_{6}$ (henceforth HCMO). The ordered double perovskites of type A$_{2}$BB$^{'}$O$_{6}$ have two sub-lattices consisting of BO$_{6}$ and B$^{'}$O$_{6}$ octahedra, which form a rock-salt type ordering.  The disorders in the form of anti-site disorder (ASD), i.e. site exchange between B/B$^{'}$ ions, are unavoidable that  might  induce competing exchange interactions  in these materials. HCMO crystallizes in a monoclinic structure with the space group symmetry $P$2$_{1}/n$, which is typically found in most of the cation-ordered double perovskite systems. Our X-ray photoelectron spectroscopy (XPS) results  reveal the exact charge state of cations in HCMO. The present double perovskite shows a long-range ferromagnetic ordering at around 76 K owing to the Co$^{2+}$-O$^{2-}$ -Mn$^{4+}$ superexchange interactions. The dc susceptibility, ac magnetization and specific heat results are consistent with such a phase transition. Moreover, the system re-entered in a glassy state (known as the re-entrant spin-glass state) below $\simeq$ 37 K, which is manifested as the frequency dependent broad peak in the ac susceptibility. 
 The observation of re-entrant glassy state could be attributed to the presence of inherent anti-site disorder and competing interactions in HCMO. Interestingly, HCMO shows large magnetocaloric effect (the corresponding isothermal magnetic entropy change is found to be $\Delta S_{\rm m}$ $\simeq$ 13.5 J/kg-K at 15 K) much below the ferromagnetic transition temperature.  Raman spectroscopy results point toward the presence of spin-phonon coupling in this novel double perovskite material. The density functional theory calculations are performed on HCMO to find the plausible ground state magnetic ordering and charge states. A detalied electronic structure analysis estabilishes the experimentally observed charge states. The ground state magnetic ordering is found to be AFM2 where the Ho spins form G-type antiferromagnetic arrangement while Co and Mn spin sublattices constitutes ferromagnetic ordering. However, no particular magnetic ordering between Ho and Co/Mn spins is observed.

\section{Experimental details}

The polycrystalline sample of Ho$_{2}$CoMnO$_{6}$ was synthesized
by conventional solid-state reaction route using high purity
initial ingredients. At first, the stoichiometric amount
of $\mathrm{Ho_{2}O_{3}}$(99.999\%, REacton), $\mathrm{CoO}$ (99.9995\%, Alfa-Aesar)
and $\mathrm{MnO_{2}}$(99.9\%, Alfa-Aesar) was mixed well. Then this
mixture was pressed into a pellet and placed in an alumina crucible and heated at $\mathrm{800^{\circ}C}$ for 24 hours, $\mathrm{900^{\circ}C}$ for 48 hours, $\mathrm{1000^{\circ}C}$ for 48 hours, and finally at $\mathrm{1250^{\circ}C}$ for 60 hours, respectively in a box furnace with several intermediate re-grinding and pelletization.

Powder X-ray diffraction (XRD) measurements were performed at room
temperature with Cu-$K_{\alpha}$ radiation ($\lambda=1.54182\mathrm{\,\mathring{A}}$)
on a Rigaku SmartLab diffractometer and was analyzed by the Rietveld method \cite{Rietveld1969} using FULLPROF software \cite{Carvajal1993}. Magnetization
measurements were carried out in the temperature range $5.0 \leq T \leq 370$
K and in the field range $0\leq H \leq 70$ kOe using a Quantum Design, SQUID VSM.
For the low-field magnetization measurements, the reset magnet mode
option of the SQUID VSM was used to remove any stray or remanent field. Specific heat measurements were performed in the temperature range 1.9 $\leq T \leq$ 200 K in zero field using the heat capacity option of the Quantum Design, PPMS. X-ray photoelectron spectroscopy (XPS) experiments were performed by using a scanning X-ray microprobe system equipped with a monochromatic aluminum $K_{\alpha}$ X-ray source (1486.6 eV) and a multi-channeltron hemispherical electron energy analyzer of PHI 5000 VERSAPROBE II, Physical Electronics system. The survey scan, as well as core-level spectra, were measured at an emission angle of 45° with a pass energy of 50 eV and 11.750 eV, respectively. The binding energy calibration was done by using C 1$s$ located at 284.6 eV. All the photoelectron measurements were performed inside the analysis chamber with an average base vacuum of 7.0 × 10$^{-10}$ mbar. A charge neutralizer was used in order to compensate for the surface charging of the samples. The total energy resolution, estimated from the width of the Fermi edge taken from a cleaned polycrystalline gold sample, was about 400 meV for monochromatic aluminum $K_{\alpha}$ line with a pass energy 11.750 eV for core level. Ar-ion sputtering has been performed at 3keV energy for 30 minutes with a raster area of 4 mm x 4 mm. The electronic structure was investigated using XPS Valence Band spectra. Raman spectra were recorded using a Jobin Yvon Horibra LABRAM-HR 800, micro-Raman spectrometer equipped with a 473 nm excitation diode laser, 1800 lines mm$^{-1}$ grating, an edge filter for Rayleigh line rejection and a charge coupled device detector giving a spectral resolution of $\sim$ 1 cm$^{-1}$
in back-scattering mode. The laser was focused to a spot size of $\sim$ 1 $\mu$m onto the flat surface of the sample using a 50$\times$ objective lens. The low temperature Raman measurements were performed employing Janis make liquid He flow type cryostat with a temperature stability of $\pm$ 0.5 K.

\begin{figure}[h]
	\centering
	\includegraphics[width=1.0\linewidth]{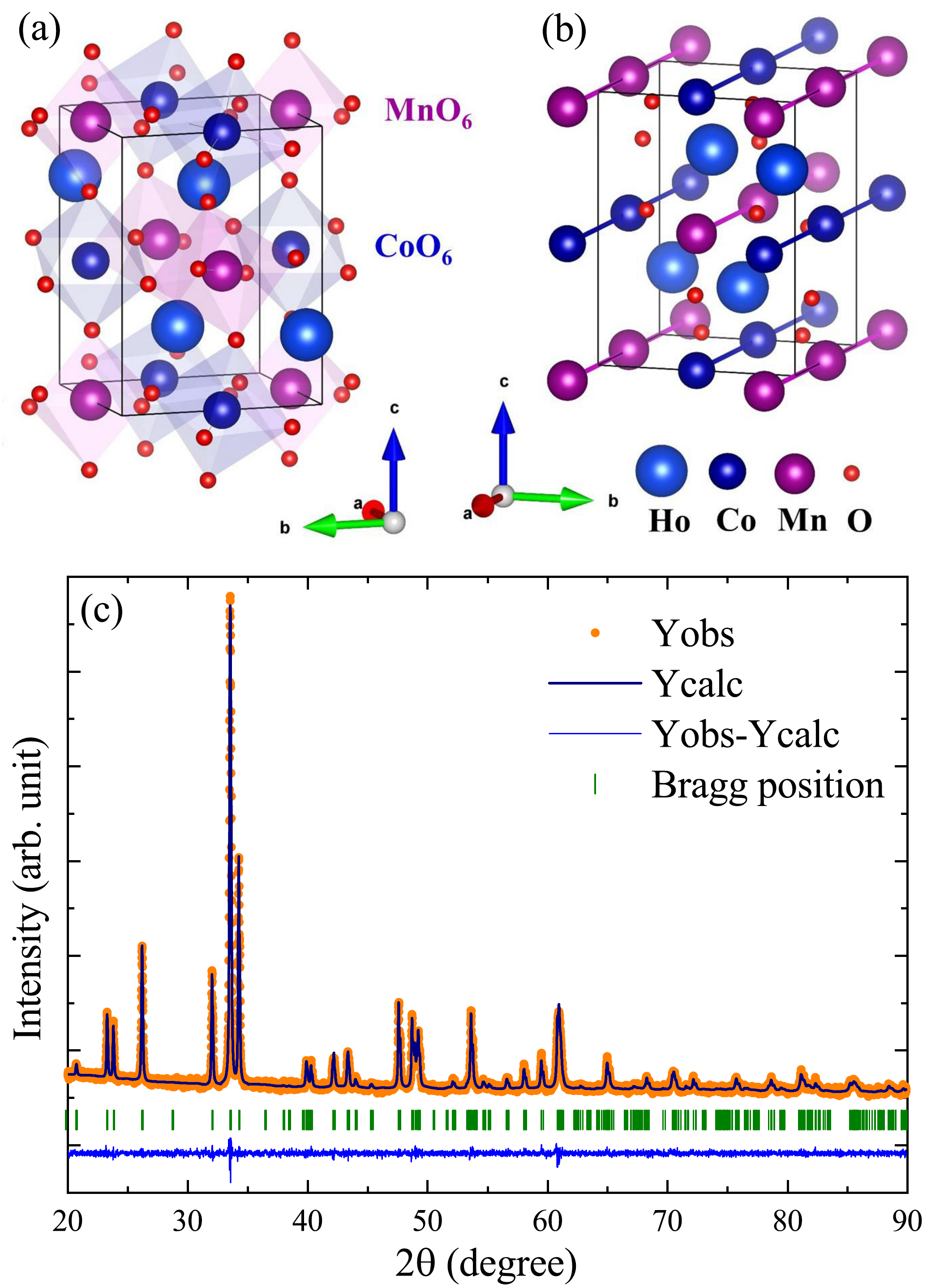}
	\caption{(a) The double perovskite crystal  structure with MnO$_{6}$ and CoO$_{6}$ octahedra in Ho$_{2}$CoMnO$_{6}$ in a unit cell.  (b) The spins in the chain are interacting along \textit{a}-axis via Co-Co and Mn-Mn chain with same bond length 5.233 \AA{}. (c) The Rietveld refinement of XRD data taken at room temperature.     }
	\label{fig:hcmochain1}
\end{figure}

\section{Results and Discussion}
\begin{center}
	\textbf{\large{}A. XRD and Crystal structure}{\large\par}
	\par\end{center}
We have refined the crystal structure of HCMO using Rietveld refinement method considering the monoclinic space group $P$2$_{1}$/$n$ with a goodness of fit $\chi^{2}$ = 1.24. The refined lattice parameters obtained from the analysis, \textit{a} = 5.2326 \AA{}, \textit{b} = 5.5869 \AA{} and \textit{c} = 7.4737\AA{} and $\beta$ = 90.12\textdegree  are in excellent agreement with the previous reports \cite{Blasco2017, Mazumdar2021}. The Rietveld refinement results are presented in Fig. \ref{fig:hcmochain1}(c) and the resulting atomic coordinates  and goodness of Rietveld refinement   are summarized in Table \ref{tab:Atomic-coordinates-of HCMO}. It is worth mentioning that although the crystal structure is derived from the $Pbnm$ perovskite, in the present A$_{2}$BB$^{'}$O$_{6}$ double perovskite material the symmetry is lower and the Co$^{2+}$ and Mn$^{4+}$ cations are distributed with a 1:1 ratio and constitute distinct B-sites. Fig. \ref{fig:hcmochain1}(a) shows the double perovskite crystal structure of Ho$_{2}$CoMnO$_{6}$ with MnO$_{6}$ and CoO$_{6}$ octahedra in a unit cell. Fig. \ref{fig:hcmochain1}(b) shows that along $a$- axis, the Co$^{2+}$ - Co$^{2+}$ and Mn$^{4+}$ - Mn$^{4+}$ ions  are constituting a uniform spin-chain separately with a minimum bond distance 5.233\AA{}. Whereas, the minimum cationic neighbors in the \textit{ab}-plane between Co$^{2+}$-Mn$^{4+}$ ions along [110] diagonals is 3.827 \AA{}  and along the $c$-axis is 3.737\AA{}, respectively. In both directions, Co-Mn pairs are forming a uniform spin-chain and the interaction along one of these directions is most likely the dominant interaction pathway in HCMO. To confirm this scenario, we also performed the electronic structure calculations using Density Functional Theory (DFT). 

\begin{table}[h]
	\caption{\label{tab:Atomic-coordinates-of HCMO}{\small{}The crystallographic
			data of Ho$_{2}$CoMnO$_{6}$ obtained from Rietveld refinement. The goodness of the Rietveld refinement as defined by the following
			parameters are $\mathrm{R_{p}}$: 2.30\%; $\mathrm{R}_{\mathrm{wp}}$:
			2.99\%; $\mathrm{R}_{\mathrm{exp}}$: 2.69\%; $\mathrm{\chi^{2}}$:
			1.24 }}
	\bigskip{}
	
	\centering{}%
\begin{tabular}{|c|c|c|c|c|c|}
	\hline
Atom	& Wyckoff position  & x & y & z & Occupancy  \\
	\hline
Ho	& 4e & 0.0203 & 0.0715 & 0.2508 & 1.000  \\
	\hline
Co	& 2d & 0.0000 & 0.5000 & 0.0000 & 1.000 \\
	\hline
Mn	& 2c & 0.5000 & 0.0000 & 0.000 & 1.000 \\
	\hline
O1	& 4e & 0.3180 & 0.3120 & 0.0550 & 1.000 \\
	\hline
O2	& 4e & 0.3000 & 0.2880 & 0.4500 & 1.000 \\
	\hline
O3	& 4e & 0.6022 & 0.9614 & 0.2380 & 1.000 \\
	\hline
\end{tabular}
\end{table}

\begin{center}
	\textbf{\large{}B. X-ray Photoelectron Spectroscopy (XPS)}{\large\par}
	\par\end{center}
X-ray photoelectron spectroscopy (XPS) is a powerful technique to probe the valence states and ligand coordination of the constituting elements in any material. In the XPS study of a material having open-shell ions, a core electron vacancy along with the open-shell gets coupled to produce the multiple structure. In addition to the main photoelectron peak features, the associated satellite peaks, chemical shifts and their relative intensities are also useful to estimate the oxidation states and ligand co-ordinations \cite{Pal2019, Pal2020}. A prior knowledge of the electronic structure of a material is essential to elucidate many of its physical properties. Hence, to understand the electronic structure of the present material HCMO, we have studied its XPS spectra. All the peak positions have been assigned following the National Institute of Standard Technology (NIST) database \cite{NISTXPS}. The deconvolution analysis of the core level XPS peaks of the relevant ions have been carried out using a combination of Lorentzian and Gaussian distribution functions.

\begin{figure}[h]
	\centering
	\includegraphics[width=0.99\linewidth]{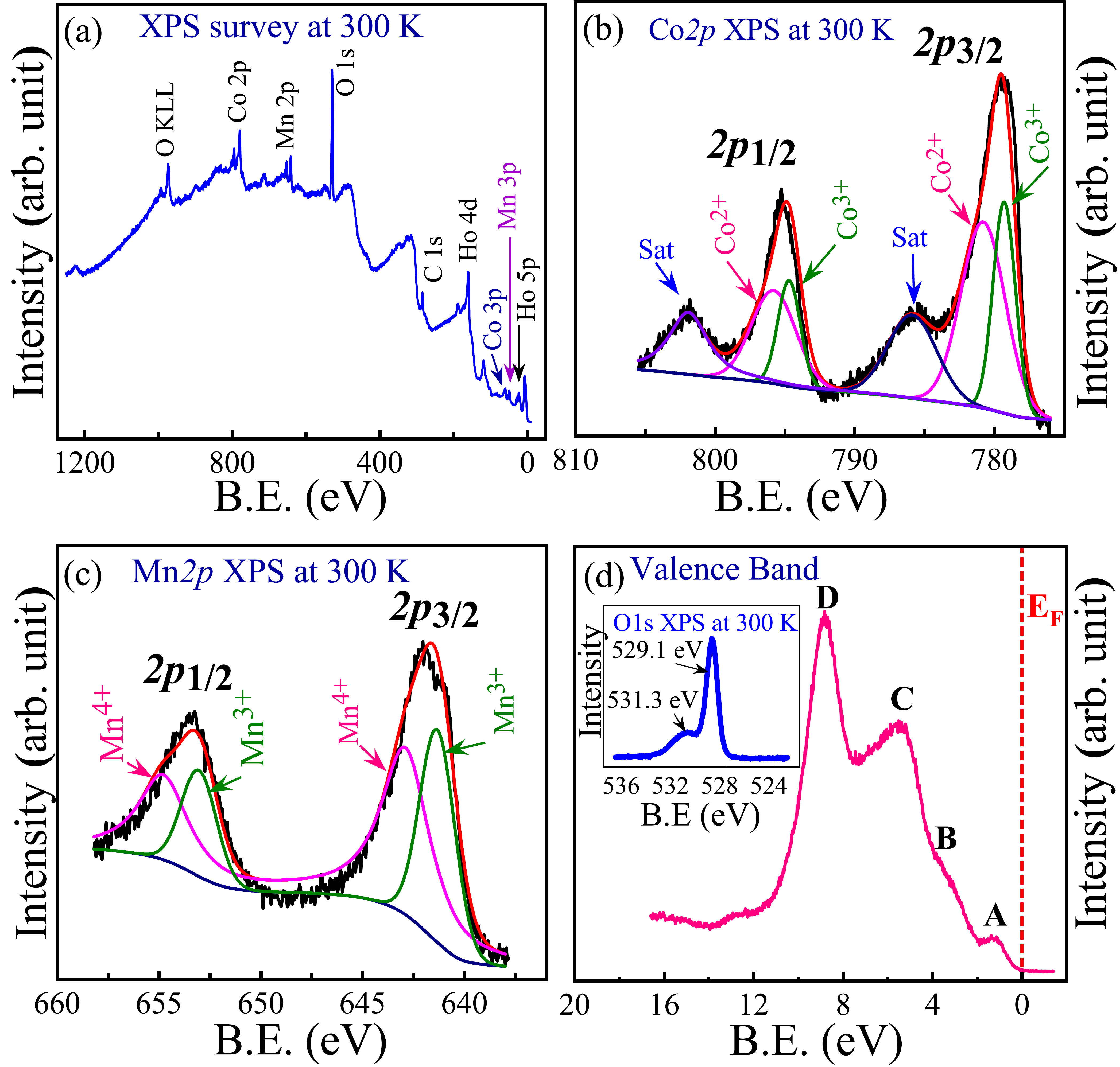}
	\caption{(a) XPS survey scan of HCMO sample at 300 K. (b) and (c) depicts the core
		level XPS spectra of Co 2$p$ and Mn 2$p$, respectively (d) demonstrates the valence
		band spectra while the inset shows the core level spectra of O 1$s$.}
	\label{fig:xps}
\end{figure}

The survey scan XPS spectra recorded at 300 K have been shown in Fig. \ref{fig:xps}(a), which confirms the presence of the elements Ho, Co, Mn, O and C in the system. The absence of extrinsic elements confirms the sample purity. The observation of the C 1$s$ peak is common and it is attributed to the extrinsic molecules absorbed from the air at the surface.

The core level Co 2$p$ XPS spectrum can provide many important information viz., spin state, nominal valence state etc. \cite{Pal2019, Pal2020, Laureti2008, Vaz2009}. Its shake-up satellite peaks are produced by poorly screened states whereas the main peaks are originated from the well-screened states. The satellite peaks in particular are very sensitive to the valence states of the cobalt ions. Fig. \ref{fig:xps}(b) demonstrates the core level Co 2$p$ XPS spectrum. It comprises of two main spin-orbit coupling peaks i.e. Co 2$p_{3/2}$ and Co 2$p_{1/2}$ positioned at the 779.43 eV and 795.1 eV, respectively. The two broad peaks situated above the main peaks are assigned as the charge transfer satellite peaks.  Eventually, such prominent satellite peaks above the main peaks are typically observed in Co 2$p$ XPS spectra of the materials containing divalent Co$^{2+}$ ions while such satellite peaks are merely absent or very weak for systems having trivalent Co$^{3+}$ ions \cite{Pal2019, Pal2020, Laureti2008}.  Hence, the observation of the satellite peaks in the present Co 2$p$ XPS spectrum indicates the presence of Co$^{2+}$ ions in HCMO. It is worth to mention here that the asymmetry and broadening of the observed peaks are indicative of the presence of mixed valence states of the Co ions \cite{Pal2020, Laureti2008, Vaz2009}. The peak positions and line-shape of the observed Co 2$p$ XPS spectrum is similar to the earlier reports showing mixed valence Co ions \cite{Pal2020}. Again, the doublet separation between the spin-orbit coupling peaks is typically found to be 15.9 eV for CoO and 15.3 eV for Co$_{2}$O$_{3}$. For the present system, the Co 2$p$ doublet separation is observed to be 15.6 eV which suggests for the existence of both Co$^{2+}$/Co$^{3+}$ ions. The Co 2$p$ XPS spectrum has been deconvoluted to estimate the concentrations of different Co ions (Co$^{2+}$/Co$^{3+}$) as shown in the Fig. \ref{fig:xps}(b). Although Co$^{2+}$ ions are found to be predominantly present in the system but the presence of Co$^{3+}$ ions is also unavoidable. This in turn can give rise to competing exchange interactions in HCMO.

The core level Mn 2$p$ XPS spectrum of the present material is depicted in Fig. \ref{fig:xps}(c). This spectrum is broadly divided into two spin-orbit coupling peaks i.e. Mn 2$p_{3/2}$ and Mn 2$p_{1/2}$, which are observed at 641.7 eV and 653.4 eV respectively.  The values of the Mn 2$p_{3/2}$ peak positions in compounds Mn$_{2}$O$_{3}$ and MnO are reported to be 641.3 eV and 642.2 eV, respectively \cite{NISTXPS}. Hence, the observed value of Mn 2$p_{3/2}$ peak position in HCMO suggests the existence of mixed valence states of Mn ions (Mn$^{3+}$/Mn$^{4+}$). For the present system, the spin-orbit splitting energy ($\Delta E $) in Mn 2$p$ XPS spectrum is found to be $\simeq$ 11.7 eV. On the other hand, $\Delta E$ for MnO$_{2}$ and Mn$_{2}$O$_{3}$ are reported to be 11.8 eV and 11.6 eV, respectively. Thus, the observed intermediate value of $\Delta E \simeq$ 11.7 eV can be presumably attributed to the mixed oxidation states of Mn ions (Mn$^{3+}$/Mn$^{4+}$) \cite{Pal2020}. The deconvolution analysis of the Mn 2$p$ XPS peaks and the concentrations of the relevant Mn$^{3+}$/Mn$^{4+}$ ions have been depicted in Fig. \ref{fig:xps}(c). The O 1$s$ core level XPS spectrum is illustrated in the inset of Fig. \ref{fig:xps}(d), which comprises of two peaks. The most intense peak at ~ 529.1 eV can be ascribed to the O$^{2-}$ ions while the smaller broad peak at ~ 531.3 eV is known to be associated with the less electron rich oxygen species (viz., O$^{2-}_{2}$, O$^{-}_{2}$ or O$^{-}$) owing to the adsorption of oxygen at the surface \cite{Pal2019, Pal2020}. 

The valence band (VB) XPS spectrum is illustrated in Fig. \ref{fig:xps}(d).  It is clear from Fig. \ref{fig:xps}(d) that no electronic states are available near the Fermi level ($E_{\rm F}$). This is a clear indication of the insulating nature of HCMO. The observed VB spectrum is made up of four peaks marked as A, B, C and D. The first spectral feature A lying immediately below Fermi energy (0-1 eV) can be attributed to the extended hybridized states of  Co 3$d$ ($e_{\rm g}$) and Mn 3$d$ ($e_{\rm g}$) \cite{Pal2019, Pal2020}. The next two features  marked as B and C  can be primarily attributed to the hybridization of states of Mn 3$d$($t_{\rm 2g}$), Co 3$d$($e_{\rm g}$) and Ho 4$f$  along with O 2$p$ states \cite{Pal2019, Pal2020}. The last peak in VB spectrum denoted as D is seemingly related to the hybridization of the extended Mn 3$d$($e_{\rm g}$), Co 3$d$($e_{\rm g}$) with O 2$p$ states while some other minor contributions owing to the  O 2$p$-Co/Mn  4$sp$ and O 2$p$-Ho  5$sd$ oxygen bonding states \cite{Pal2019, Pal2020}.

\begin{figure}[h]
	\centering
	\includegraphics[width=0.82\linewidth]{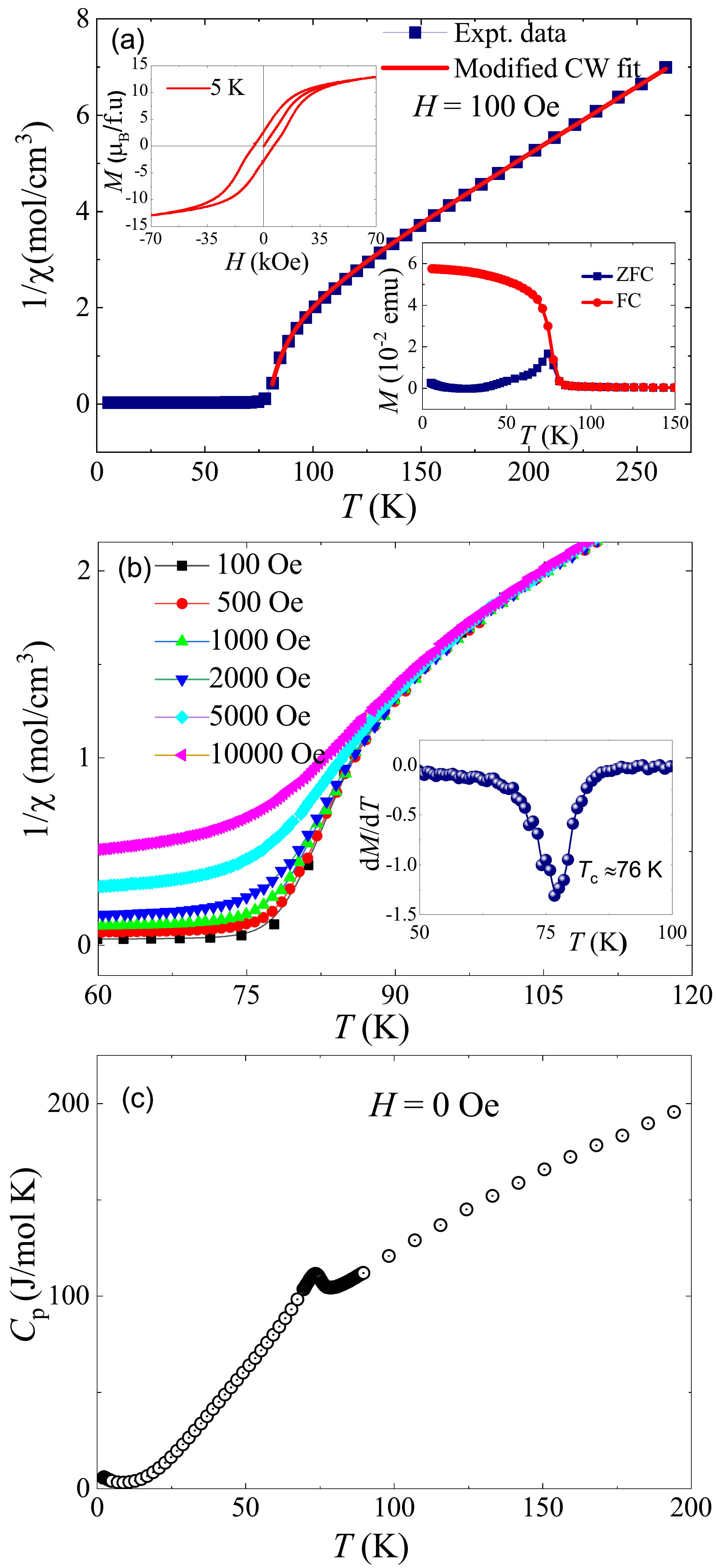}
	\caption{(a) The temperature dependence of inverse magnetic susceptibility in an applied magnetic field of 100 Oe. The upper inset depicts  $M(H)$ loop at 5 K reflecting FM order in HCMO. The lower inset shows a closer view of the temperature dependence of magnetization recorded in ZFC-FC modes in 100 Oe, which shows a bifurcation at $T$ = 76 K. (b) The temperature dependence of 1/$\chi$ in different magnetic fields. The inset shows the derivative of magnetization  (d$M$/d$T$) vs. $T$ taken in FC mode in  an applied magnetic field of 500 Oe clearly depicting an anomaly at 76 K. (c) Temperature dependence of specific heat of HCMO in zero field. A peak around 73.4 K is due to the long range ordering.} 
	\label{fig:chiinversemodifiedcw-fit}
\end{figure}

\begin{center}
	\textbf{\large{}C. Magnetization and specific heat}{\large\par}
	\par\end{center}
In HCMO, one expects competing exchange interactions due to the presence three different magnetic ions with different energy scale of interactions  in the spin-lattice with inevitable disorder. The competing interactions may  drive the system towards a unconventional magnetic state. In order to probe the ground state magnetic proprieties of this material,  we measured the dc magnetization $\mathit{M}(T)$ as a function of temperature on a hard pellet of HCMO in zero field cooled (ZFC) and field cooled (FC) modes in several magnetic fields ($\mathit{H}$) using a Quantum Design, SVSM. The magnetic susceptibility data show a paramagnetic behavior at high temperature and an anomaly around $\sim$ 76 K suggesting the presence of a  magnetic phase transition (in Fig. \ref{fig:chiinversemodifiedcw-fit}(a)). 

Similarly to the previous magnetization study on  HCMO \cite{Mazumdar2021}, the susceptibility data were fitted with the Curie-Weiss (CW) law given by Eq. $ \chi^{-1}= \left( \chi_{0}+\frac{C_{RE}}{T-\theta_{RE}}+\frac{C_{TM}}{T-\theta_{TM}}\right) ^{-1}$, where $\chi_{0}$, $C_{\rm RE}$/$C_{\rm TM}$ and $\theta_{\rm RE}$/$\theta_{\rm TM}$ are temperature independent susceptibility (arising from the core diamagnetic and paramagnetic Van-Vleck contributions), Curie constant and the Curie-Weiss temperatures for rare-earth and transition metal ions, respectively. In HCMO, the magnetic properties are governed by both rare-earth  Ho$^{3+}$ (4$f^{10}$, $^{5}I_{8}$; $J$ = 8) ions  and transition metal ions (Co$^{2+}$ and Mn$^{4+}$). The main features of our observations from the magnetization measurement are described below. The material shows a long range FM ordering around 76\,K due to the superexchange interaction between Co$^{2+}$ ($t_{\rm 2g}^{5}$, $e_{\rm g}^{2}$)  and Mn$^{4+}$ ($t_{\rm 2g}^{3}$, $e_{\rm g}^{0}$) magnetic moments.  From the CW fit, we obtained $\theta_{\rm RE}$ = - 4.2\,K, $\theta_{\rm TM}$ = 79.4\,K. The small and negative value of $\theta_{\rm RE}$ suggests the presence of a weak antiferromagnetic interaction between  Ho$^{3+}$ moments and the sub-lattice consisting of transition metal ions Co$^{2+}$/Mn$^{4+}$ or associated with the crystal electric field excitations of $^{5}I_{8}$  multiplet of the  Ho$^{3+}$ ion. Ho$^{3+}$  moments may undergo a phase transition at  lower temperature  below 2 K. While the positive value of $\theta_{\rm TM}$ indicates a ferromagnetic interaction between Co$^{2+}$ and Mn$^{4+}$ in the host spin-lattice. The presence of AFM  and FM interactions suggest that HCMO hosts a frustrated spin-lattice. The obtained effective magnetic moment ($\sqrt{8C} \mu_{\rm B}$) for transition metal cations is 5.44\,$\mu_{\rm B}$. This value of $\mu_{\rm eff}$ is close to that of the theoretically expected one ($\mu_{\rm eff}= \sqrt{\mu^{2}_{\rm eff}( Co^{2+}) + \mu^{2}_{\rm eff}(Mn^{4+})}$ for the Co$^{2+}$ (3$d^{7}$, $S$ = 3/2) and Mn$^{4+}$ (3$d^{3}$, $S$ = 3/2) sublattice system [($\mu_{\rm eff}$)$_{\rm theo}$ = 5.47\,$\mu_{\rm B}$/ f.u.], indicating the presence of divalent Co and tetravalent Mn sub-lattices, respectively \cite{Das2019}. Whereas the effective moment for rare-earth ion Ho$^{3+}$ in HCMO is 11.03 $\mu_{\rm B}$, which is a bit larger than that expected for free Ho$^{3+}$ ion but is consistent with previous reports \cite{Mazumdar2021}. A bit larger value of Ho$^{3+}$  magnetic moment compared to free ion value possibly indicating a weak interaction between Ho$^{3+}$ and Co/Mn sublattices. Fig. \ref{fig:chiinversemodifiedcw-fit}(b) shows the temperature dependence of inverse susceptibility data in different applied magnetic fields displaying  a phase transition around 76 K. The inset of Fig. \ref{fig:chiinversemodifiedcw-fit}(b) shows the derivative of magnetization  (d$M$/d$T$) vs. $T$  taken in FC mode in 500 Oe clearly depicting an anomaly at 76 K owing to FM order.  Furthermore, the magnetic hysteresis behavior of magnetization $M(H)$ at 5 K (see the upper inset of Fig.  \ref{fig:chiinversemodifiedcw-fit}(a)) confirms the presence of FM magnetic order. The behavior of  inverse magnetic susceptibility above $T_{\rm c}$ is most likely due to the persistence of short-range spin correlations in this double perovskite.

The bifurcation of ZFC-FC data below  $T_{\rm c}$ = 76 K in an applied field of 100 Oe as shown in the inset of Fig. \ref{fig:chiinversemodifiedcw-fit}(a), indicates the presence of a spin-glass \cite{Binder1986} state in HCMO. The FC magnetization increases upon lowering the temperature below the transition temperature and tends to saturate at low temperature indicating the polarization of 4$f$ ( Ho$^{3+}$)  moments in an applied magnetic field. In order to gain further insights into the  magnetic ordering and the low-energy excitations in HCMO, we measured the specific heat of HCMO at constant pressure $\mathit{C}_{\rm P}(T)$ in the $\mathit{T}$- range $1.9 \leq T \leq 200$ K in zero field. The sharp anomaly at around 73.4 K in the $C_{\rm p}(T)$  vs $T$ data supports the presence of a long-range  FM ordering as observed in the dc magnetization.  However, there is a little discrepancy between the $T_{\rm c}$ observed in magnetization and specific heat experiments. A bit higher value of $T_{\rm c}$ in magnetization  is most likely due to  the fact that the transition temperature, $T_{\rm c}$ is quite sensitive to magnetic field used for magnetization measurements, while the specific heat was measured in zero field. Another plausible scenario regarding the slight discrepancy in $T_{\rm c}$ observed by  magnetization and specific heat measurements could be due to the fact that dissimilar experimental techniques track a bit different spin dynamics. This small difference in $T_{\rm c}$ measured by dissimilar experimental  techniques  is  also observed in other frustrated magnets (see for instance reference \cite{Kumar2019a}), however, the exact origin of which is not clear at present. The enhancement of specific heat below 5 K (see Fig. \ref{fig:chiinversemodifiedcw-fit}(c)) suggests the onset of a phase transition at much lower temperature owing to  the presence of a weak exchange interaction between Ho$^{3+}$ moments.

\begin{center}
	\textbf{\large{}D. AC Susceptibility}{\large\par}
	\par\end{center}
\begin{figure}[h]
	\centering
	\includegraphics[width=0.97\linewidth]{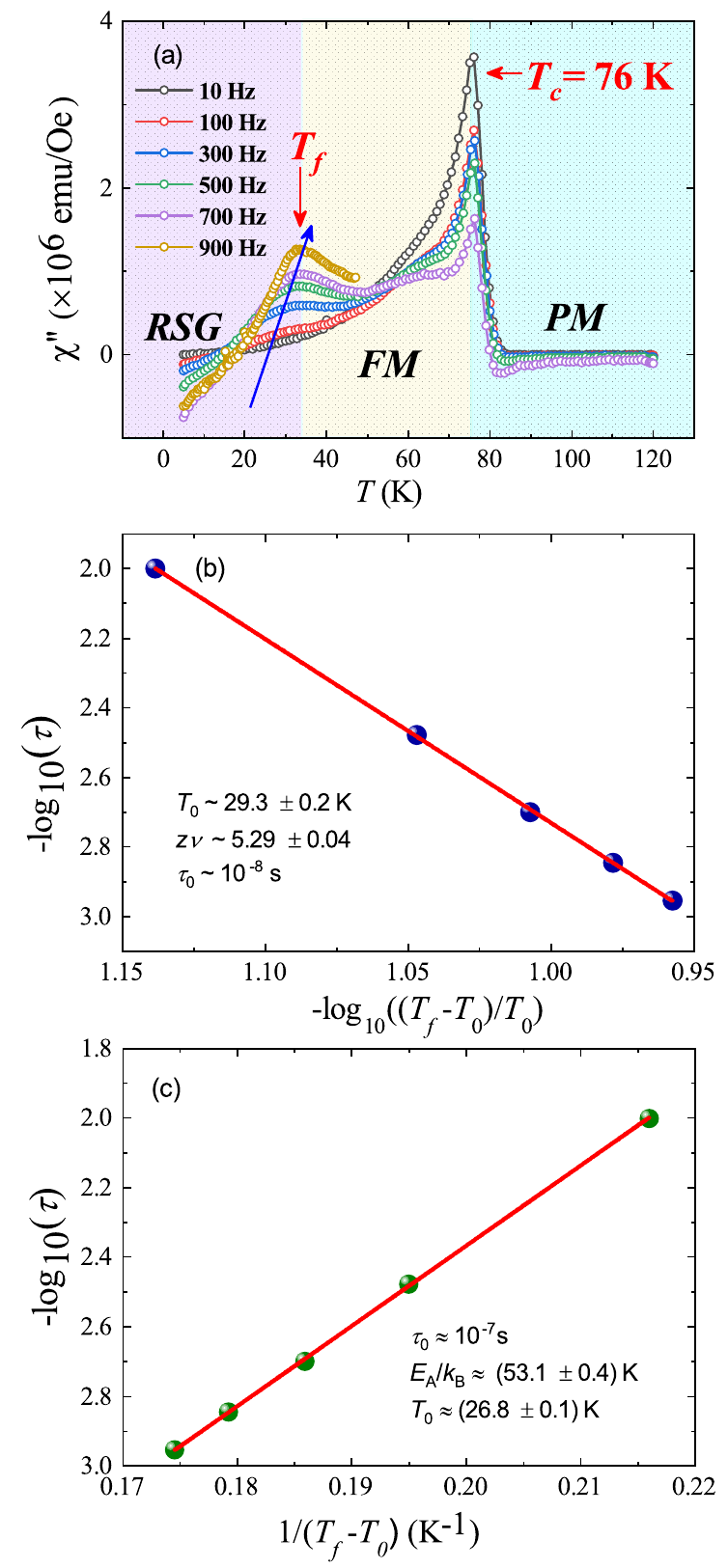}
	\caption{(a) The temperature dependence of the imaginary part of the ac susceptibility $\chi^{''}$ data at different frequencies with an applied ac field of 3.5 Oe. The inverse of frequency vs. freezing temperature ($T_{\rm f}$) data is fitted with (b) the critical slowing down model and (c) with the Vogel-Fulcher law.}
	\label{fig:chiacimgall}
\end{figure}

The splitting of ZFC and FC curves below $\sim$ 76 K suggests that a spin-freezing mechanism is at play which could be associated with anti-site disorder and frustration in HCMO. AC susceptibility  is a very sensitive probe to track spin freezing mechanism and shed insights into the characteristics of magnetization dynamics  in the ordered state  of the frustrated magnets. Generally, the spin dynamics slow down below the freezing temperature  $T_{\rm f}$ owing to the cluster formation of spin domains. In order to confirm the  dc susceptibility results and  understand glassy behavior of this frustrated magnet, we performed ac susceptibility measurements at several frequencies. Fig. \ref{fig:chiacimgall}(a) depicts the imaginary part of the ac susceptibility [$\chi^{''}(T)$] which shows a sharp peak at the FM ordering temperature $T_{\rm c}$ = 76 K and a broad  maximum around  $T_{\rm f}$ = 30 K. The maximum at $T_{\rm f}$ = 30 K shifts towards higher temperature upon increasing the frequency indicating a glassy spin dynamics leading to a slow relaxation of magnetization.  To investigate the variation of $T_{\rm f}$ with frequency, we have analyzed it following the scaling law relevant for critical dynamics \cite{Djurberg1997,Mukadam2005} given by

\begin{equation}
	\tau = \tau_{0} \left( \frac{T_{\rm f}-T_{0}}{T_{0}}\right)^{-zv} 
	\label{CDS}
\end{equation}

Here, $T_{\rm f}$ is the freezing temperature where $f$ stands for the the corresponding frequency for which $\chi^{''}(T)$ attains a maximum, $T_{0}$ is the equivalent spin-glass freezing temperature when $f$\textrightarrow\,0 Hz
and $H_{DC}$ \textrightarrow\,0 Oe, and $\tau_{0}$ is related to the characteristic spin flipping time ($f_{0} = \frac{1}{\tau_{0}}$); $zv$ is the dynamical critical exponent. Fig. \ref{fig:chiacimgall}(b) shows the critical dynamic scaling fit. The best fitting yields $\tau_{0}$ = 10$^{-8}$ s, $T_{0}$ = (29.3 $\pm$ 0.2) K, and $zv$ = (5.29 $\pm$ 0.04), which is in reasonable  agreement with that expected for a cluster spin-glass state (4 $<$ $zv$ $<$ 12). For a canonical spin-glass system, $\tau_{0}$ lies between 10$^{-12}$ s and 10$^{-13}$ s, which is smaller than the observed value
of 10$^{-8}$ s by a few orders. The longer relaxation time $\tau_{0}$ indicates that the spin-freezing characteristics is of cluster-glass type rather than atomic.

To explore further the characteristics of inter cluster interactions, well-known Vogel-Fulcher (VF) model (see Eq.\ref{VF}) was employed for fitting the relaxation time $\tau$ vs. $T_{\rm f}$ curve.  Fig. \ref{fig:chiacimgall}(c) shows the corresponding VF model fit. 

\begin{equation}
	\tau = \tau_{0}\; {\rm exp} \left( \frac{E_{\rm A}}{k_{B}(T_{\rm f}-T_{0})}\right) 
	\label{VF}
\end{equation}

where $T_{0}$ represents inter-cluster interaction strength and $E_{\rm A}$ is the activation energy required to overcome the barrier of the metastability of the spin-glass (SG) state. The best fitting yielded for $\tau_{0}$ = 10$^{-7}$\,s, $T_{0}$
= (26.8 $\pm$ 0.1)\,K, and $E_{\rm A}$/$k_{\rm B}$ = (53.1 $\pm$ 0.4)\,K.  The large value of $\tau_{0}$ is expected from interacting magnetic spin clusters. All the features observed from ac magnetization results point towards the stabilization of a re-entrant spin-glass (RSG) ground state below $T_{\rm f}$ = 30\,K.

\begin{figure}[h]
	\centering
	\includegraphics[width=0.90\linewidth]{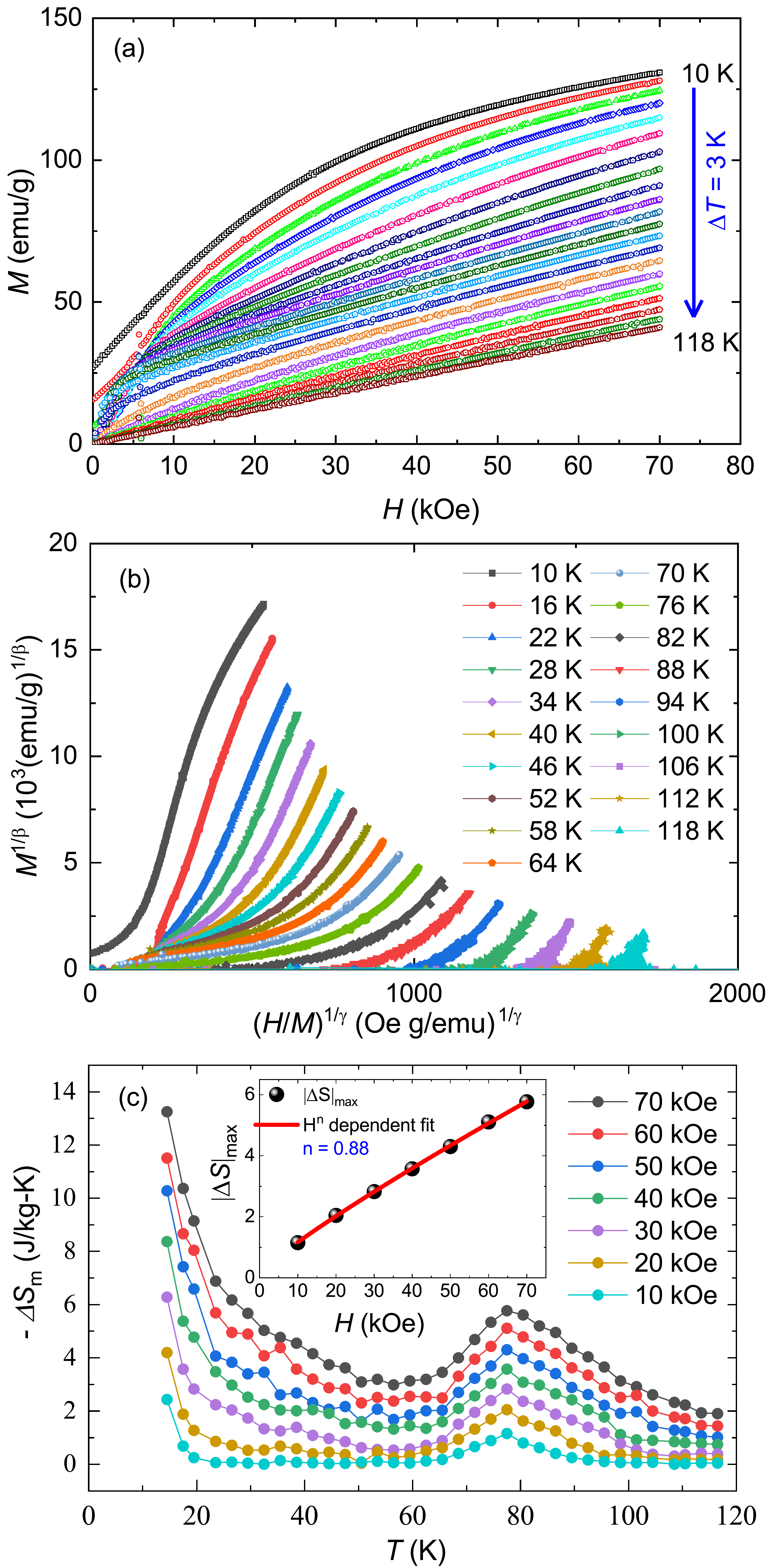}
	\caption{(a) Magnetization isotherms at several temperatures.(b) The modified Arrott plot of isotherm magnetization following mean field approximation. (c) The magnetic entropy change - $\Delta S_{m}$ as a function of temperature showing a peak at FM ordering temperature  with $T_{c}$ = 76 K. Inset shows the maximum value of magnetic entropy change $|\Delta S_{m}|_{max}$ vs. $H$ with a power law fit as represented by a solid red line. }
	\label{fig:MH_Magcal}
\end{figure}

We also measured the isotherm magnetization at several temperatures and Fig. \ref{fig:MH_Magcal}(a) shows the corresponding plot. As depicted in Fig. \ref{fig:MH_Magcal}(a),  the magnetization increases upon increasing the external magnetic field and does not saturate up to 7 T. Using the magnetic isotherm data around $T_{\rm c}$, we have plotted the modified Arrott plot \cite{Arrott1957,Yeung1986} considering mean field model (where the critical exponents are $\beta$ = 0.5 and $\gamma$ = 1), which corroborates the presence of ferromagnetic order in this material. Fig. \ref{fig:MH_Magcal}(b) represents the corresponding Arrott plot with a set of parallel lines near $T_{\rm c}$ = 76 K in high field. The non-linear behavior of  Arrott curves suggests the presence of inhomogeneous magnetism possibly related to the fact that the  unavoidable disorder strongly affects the underlying spin dynamics of HCMO \cite{Herzer1980}.

External perturbations such as magnetic field play a crucial role in modifying the  ground state properties and internal energy of double perovskite based  magnetic material under study \cite{GschneidnerJr2005}. The temperature variation in magnetic materials  leading to change in entropy due to an adiabatic change of external magnetic field is known as magnetocaloric effect relevant for magnetic refrigeration \cite{Provenzano2004, Tegus2002a}.  Geometric frustration plays an important role in the enhancement of the change in magnetic entropy in the presence of a  magnetic field, which in turn  can lead to a large magnetocaloric effect in double perovskites \cite{Zhitomirsky2003}. In order to extract the temperature dependence of the change in magnetic entropy in HCMO, we have measured magnetization isotherms at several temperatures. For the characterization of the magnetocaloric response of a material, three main parameters can be studied: the isothermal magnetic entropy change, $\Delta S_{\rm m}$; the adiabatic temperature change, $\Delta T_{\rm ad}$; and the refrigerant capacity. Generally,  $\Delta S_{\rm m}$ can be calculated indirectly from the experimental magnetization curves using Maxwell's relation.
\begin{equation}
	\Delta S_{\rm m} = \mu_{\rm 0} \int^{H}_{0} \left(\frac{\delta M}{\delta T}\right)_{H}\delta H
	\label{Maxwell relation}
\end{equation}
  
The resulting temperature dependence of the magnetic entropy change $\Delta S_{\rm m}$  in different applied magnetic fields is shown in Fig. \ref{fig:MH_Magcal}(c). In ferromagnetic materials, the magnetic field aligns the moments along the field direction by overcoming the thermal fluctuations, leading to a decrease of the magnetic entropy change,  $\Delta S_{\rm m}$ < 0. Whereas, the external magnetic field plays a vital role in  rotating the spins against their preferential  directions yielding an increase of magnetic entropy, $\Delta S_{\rm m}$ > 0 in antiferromagnetic materials \cite{Phan2007,Lampen2014}.
As depicted in Fig.  \ref{fig:MH_Magcal}(c), a peak in -$\Delta S_{\rm m}$ around $T_{\rm C}$ corroborates a  phase transition from a paramagnetic state to a ferromagnetically ordered state. The value of $\Delta S_{\rm m}$ increases upon lowering the temperature and attains a maximum value $\sim$ 13.5 J/kg-K at low temperature in a magnetic field of 7 T. The enhancement of -$\Delta S_{\rm m}$ at low temperature  indicates the onset of a magnetic ordering most likely due to the presence of a weak exchange interaction between  Ho$^{3+}$ moments. As  presented in Fig. \ref{fig:MH_Magcal}(c), HCMO shows a large magnetocaloric effect (the corresponding isothermal magnetic entropy change is found to be $\simeq$ 13.5 J/kg-K at around 15 K) much below the ferromagnetic transition temperature \cite{Mazumdar2021}. The obtained large value of magnetic entropy change owing to adiabatic demagnetization is quite unusual in double perovskites, which  suggests that HCMO is a promising candidate for magnetic refrigeration technology. 
Furthermore, the variation of the maximum value of entropy change i.e. ($|\Delta S_{\rm m}|_{\rm max}$) with the magnetic field obeys a power law ($\propto H^{n}$) reflecting a second order phase transition, where $n$ is a temperature dependent parameter \cite{Oesterreicher1984}.  As shown in the inset of Fig. \ref{fig:MH_Magcal}(c), the power law fit  yields a magnetic-ordering parameter $n$ = 0.881, which is close to that of the value obtained in another double perovskite Dy$_{2}$CoMnO$_{6}$ \cite{Li2017}. The  value of the exponent $n$ which is related to the the critical exponents ($\beta$ and $\gamma$) near the $T_{\rm C}$ and it is defined as $n$ = 1+ $\frac{(\beta-1)}{(\beta+\gamma)}$ \cite{Singh2020a}. The obtained value of the parameter $n$ indicates that the magnetic order in HCMO might be best described by modified Arrott plots viz. - 3D-Heisenberg model, the 3D-Ising model, or tri-critical model than the mean-field framework, which invokes further studies.

\begin{figure}[h]
	\centering
	\includegraphics[width=0.99\linewidth]{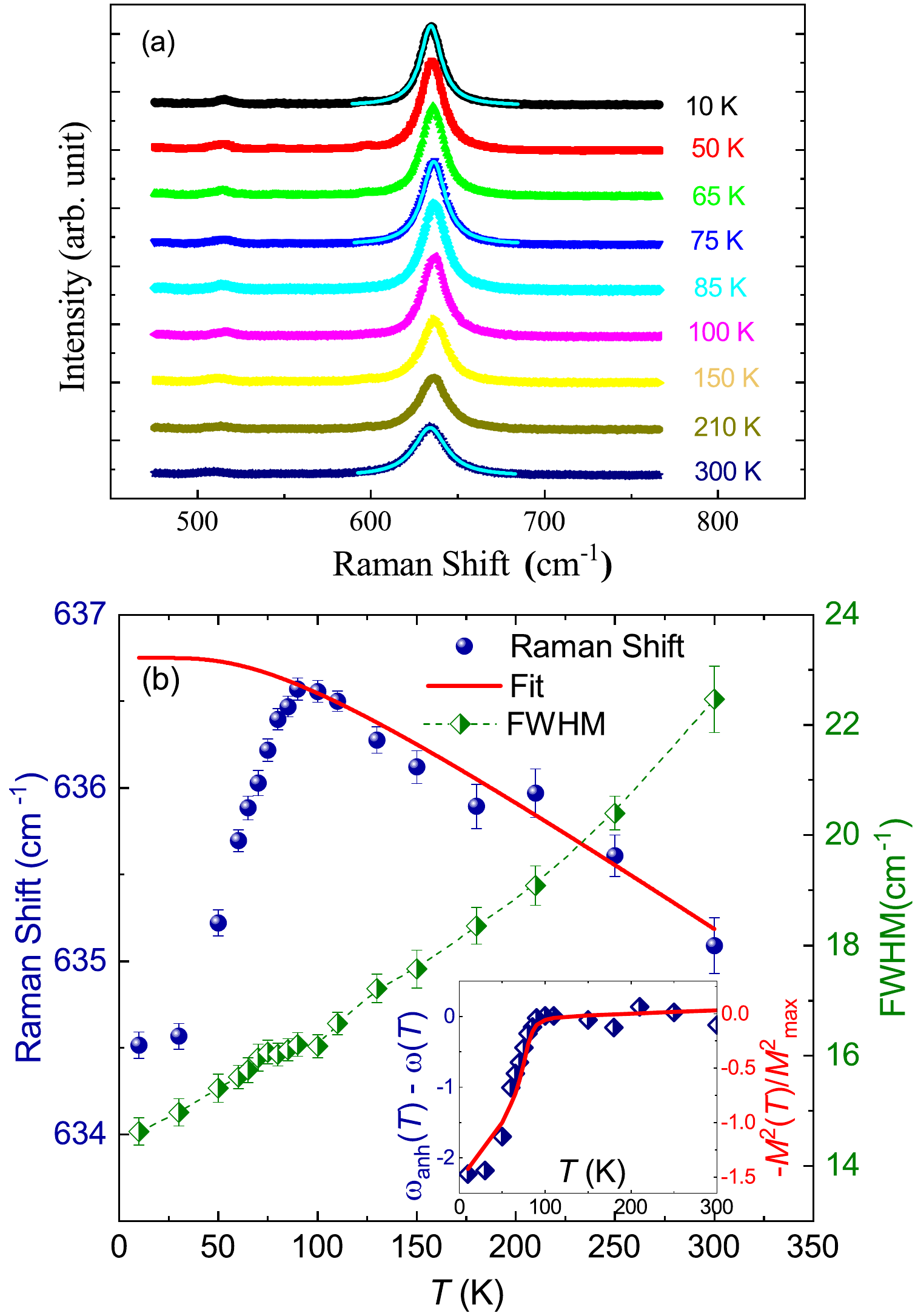}
	\caption{(a) Raman spectra with Lorentzian fit at a few selected temperatures. (b) The anharmonic nature of the prominent Raman band (635 cm$^{-1}$) as a function of temperature. Solid red line represents the standard $\omega_{anh}(T)$ dependence expected in the absence of spin-phonon coupling. The variation of the full width at half maximum (FWHM) with temperatures shown on right $y$-axis. The inset shows the departure from the anharmonic behavior $\Delta \omega(T)$ of the prominent Raman band is commensurate with the normalized magnetic moment $M^{2}(T)$/$M^{2}_{max}$ where temperature is an implicit parameter.
	}
	\label{fig:raman}
\end{figure}

\begin{center}
	\textbf{\large{}E. Raman Spectra}{\large\par}
\par\end{center}
The complex interplay between emergent degrees of freedom can lead to interesting magnetic phenomena in double perovskites. Raman spectroscopy is an ideal probe to track competing magnetic
ordering phenomena and exotic collective excitations in correlated quantum magnets. This technique has been very successful in revealing interesting insights into the phonon characteristics and spin–phonon coupling of many interesting perovskite based oxide materials \cite{Granado1999,Laverdiere2006, Iliev2007}. In order to understand the coupling between spin and phonon in HCMO, we measured Raman spectra down to 10 K. The Raman spectra are presented in Fig. \ref*{fig:raman}(a) which shows the most prominent Raman band at 635 cm$^{-1}$ and another less intense band at 515 cm$^{-1}$. As the less intense Raman band at 515 cm$^{-1}$ is not shifting that much with temperature, we focused our data at 635 cm$^{-1}$ band. We have fitted each Raman spectra with a Lorentzian peak profile (shown for a few selected temperatures) to determine the exact peak position and the full width at half maxima as well. Fig. \refeq{fig:raman} (b) shows the plot of the temperature dependence of Raman shift. The Raman shift increases upon decreasing the temperature and this shift suddenly drops below 85 K, which indicates the dominant role of ferromagnetic ordering on phonon frequencies in this material. Such temperature dependence of Raman shift could be associated with (i) anharmonic frequency shift at constant volume,  as proposed by Balkanski \cite{Balkanski1983} in the absence of structural
phase transitions, and (ii) spin–phonon coupling. The anharmonicity of phonon modes can be represented by the following relation

\begin{equation}
				\omega_{\rm anh}(T)= \omega_{0}-C[1+2/(e^{\hbar \omega/ k_{B}T}-1)]
\end{equation}
\begin{equation}
	\Delta \omega(T)= \omega_{\rm anh}(T)-\omega_{0} \propto \frac{M^{2}(T)}{M^{2}_{\rm max}}
	\label{spin-spin}
\end{equation}
where, $\omega_{0}$ and $C$ are adjustable parameters. At high temperature, the Raman shift of the most intense Raman band at 635 cm$^{-1}$ follows the anharmonic behavior as shown in Fig. \ref*{fig:raman}(b). A noticeable deviation of the Raman shift around 85 K suggests the softening of phonon modes associated with spin-phonon interactions possibly due to stretching of (Co/Mn)O$_{6}$ octahedra in HCMO. Such behavior of the Raman shift around 85 K is known to occur due to the spin-phonon coupling ascribed to the role of lattice vibrations in tuning the exchange integral in a magnetic material. The onset of softening of phonon modes around 85 K indicates the existence  of short range ordering which is consistent with our magnetization results. It is worth to mentioning that this type of magnetic-order-induced renormalization  of phonon frequencies \cite{Granado1999,Laverdiere2006} has been observed in several double perovskites such as La$_{2}$CoMnO$_{6}$ \cite{Iliev2007} Ba$_{2}$NiMnO$_{6}$ \cite{Iliev2008}. Understanding the role of competing magnetic order and spin correlations in tuning the phonon frequencies in double perovskite is an interesting setting in the context of novel material design.  As per the phenomenological model employed in Refs. \cite{Granado1999,Laverdiere2006}, the effect in the phonon of a renormalization of electronic states in the proximity of FM order is proportional to the spin-spin correlation of the localized nearest neighbor spins. In the mean field approximation, the spin-spin correlation between Co$^{2+}$ and Mn$^{4+}$ spins is proportional to M$^{2}$(T)/M$^{2}_{\rm max}$ in the present double perovskite, where $M(T)$ represents the average magnetization per magnetic ion at a temperature $T$.  Below 85 K, $\Delta \omega(T)$ closely follows relation \ref{spin-spin} which is shown within the inset of Fig. \ref {fig:raman}(b). Our Raman results rule out the presence of a structural phase transition as evidenced by the absence of remarkable features in the Raman spectra of HCMO in the temperature range of investigation, which is consistent with specific heat  and previous neutron scattering studies \cite{Blasco2017}. The existence of spin-phonon coupling suggests  that there is an innate connection between magnetism and lattice in this 3$d$-4$f$ double perovskite.
\begin{table*}
	\caption{The relative energy differences ($E_{\rm rel}$) in meV/f.u. with respect to the most stable magnetic configuration and corresponding local spin magnetic moments at Ho, Co and Mn in $\mu_{\rm B}$. The unbracketed and bracketed terms in $E_{\rm rel}$ represents the relative energies without and with considering $U_{\rm eff}$ on Mn.}
	\label{DOS}
	\scalebox{0.98}{
		\begin{tabular}{|c|c|c|c|c|c|c|c|c|c|c|c|c|c|c|c|c|c|c||c|c|}
			\hline
			\multirow{3}{*}{Config} & \multicolumn{4}{c|}{GGA} & 
			\multicolumn{12}{c|}{GGA+U} \\
			\cline{2-17}
			& \multicolumn{4}{c|}{$U$ = 0} & \multicolumn{4}{c|}{$U_{\rm eff}$ = 2eV} & \multicolumn{4}{c|}{$U_{\rm eff}$ = 4eV} & \multicolumn{4}{c|}{$U_{\rm eff}$ = 6eV} \\
			\cline{2-17}
			& $E_{\rm rel}$ & Ho&Co&Mn & $E_{\rm rel}$ & Ho&Co&Mn & $E_{\rm rel}$ & Ho&Co&Mn & $E_{\rm rel}$ & Ho&Co&Mn\\
			\hline
			FM&0 &3.81 & 2.30 & 2.90 & 18.15(8)& 3.92&2.45 & 2.90 &16.27(6.25) &4&2.57 &2.88 &13.7(4.37)&4.15&2.78&2.81\\
			\hline
			AFM1&72.73.4 & 3.80& 2.20&2.94&39.84(28.55)& 3.90&2.50 &2.80 & 35.42(15.27)&3.98&2.70&2.7&33.19(7.52)&4.05&2.85&2.83\\
			\hline
			AFM2&72.87 &3.82 &2.22 &2.92 & 0&3.92 & 2.45& 2.92&0&4&2.58 & 2.88&0&4.1&2.79&2.82\\
			\hline
			AFM3& 210.2&3.82 &2.40 &2.82 & 117.04(220.25)&3.92 &2.52 &2.9 & 96.45(289.2)&4&2.62 & 2.85&64.8(345.7)&4.13&2.81&2.85\\
			\hline
			AFM4&83.56 &3.8 &2.35 & 2.94&14.43(10.58) &3.92 & 2.43&2.91 &21.08(13.55) &4& 2.6& 2.88&26.28(15.81)&4.13&2.78&2.82\\
			\hline
	\end{tabular}}
\end{table*}

\begin{figure*}
	\centering
	\includegraphics[width=0.98\linewidth]{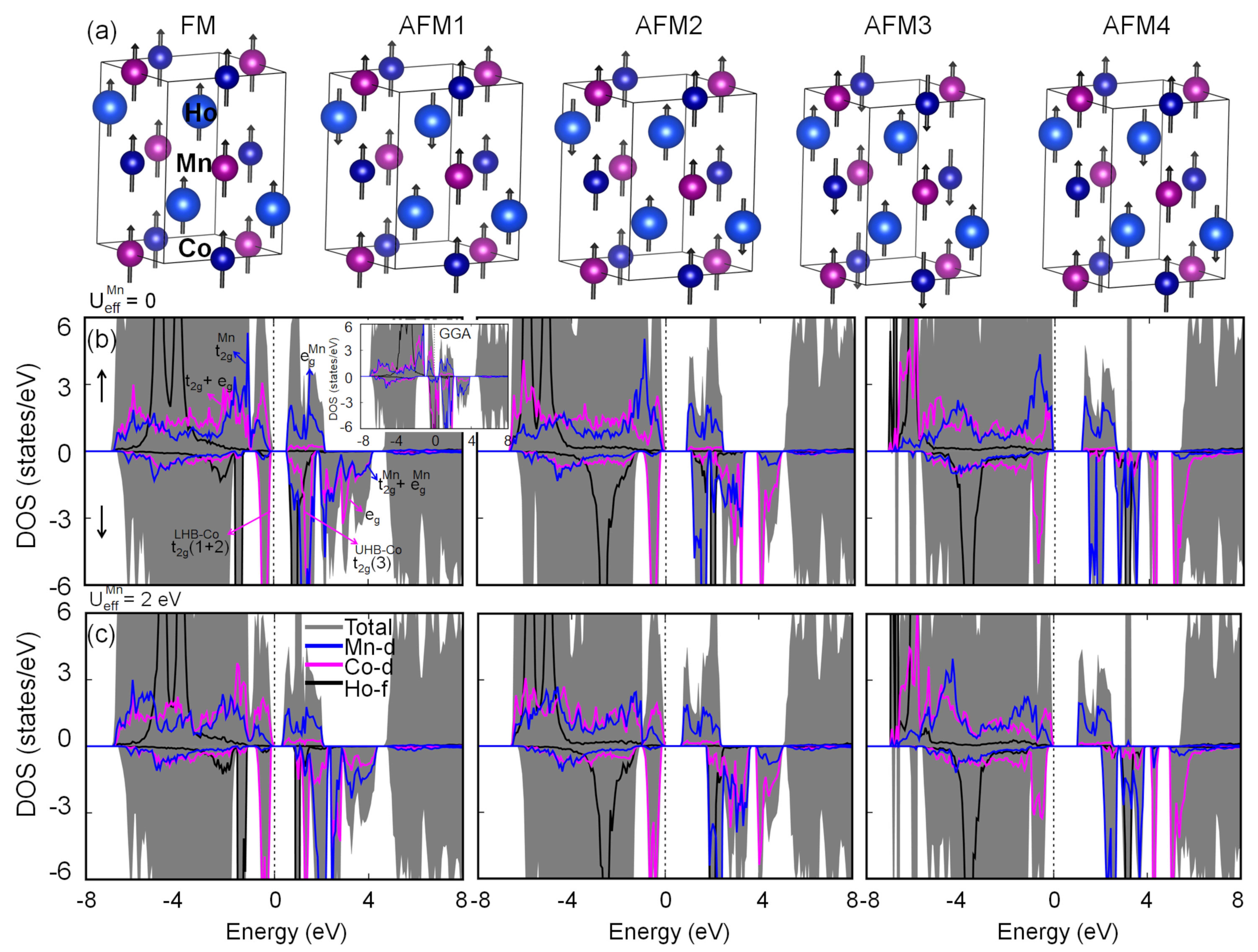}
	\caption{Panel (a) The five considered magnetic arrangements, namely, ferromagnetic (FM), antiferromagnetic AFM1, AFM2, AFM3 and AFM4, respectively. The AFM1, AFM2 and AFM4 configurations are designed in such a way that Ho atoms form A-type, G-type and C-type antiferromagnetic arrangements whereas in the AFM3 configuration the Co and Mn atoms form G-type arrangements while keeping ferromagnetic interaction between Ho atoms. Panel (b) The spin and atom resolved DOS of most stable magnetic configuration AFM2 as a function of $U_{\rm eff}$. The $U_{\rm eff}$ is varied uniformly on Ho and Co ($U_{\rm eff}$ = 2, 4 and 6 eV) atoms while keeping zero for Mn. The inset depicts the DOS for the GGA case. Panel (c) The variation in DOS by keeping $U_{\rm eff}$ fixed on Mn while varying on Ho and Co.}
	\label{bulk_bands_evo}
\end{figure*}
\begin{center}
	\textbf{\large{}F. Electronic and Magnetic structure of HCMO from Density Functional Theory (DFT) calculations}{\large\par}
	\par\end{center}
In order to gain further insights into the experimentally observed charge and magnetic configurations, density functional theory (DFT) calculations were carried out to find the ground state electronic and magnetic properties of HCMO. The calculations were carried out on five magnetic configurations namely, FM, AFM1, AFM2, AFM3 and AFM4 (see Fig. \ref{bulk_bands_evo}). For this purpose, we considered experimentally obtained crystal structure and used the projector augmented wave (PAW) \cite{Kresse1999} method as implemented in the Vienna ab-initio simulation package (VASP) \cite{Kresse1996}. The generalized gradient approximation (GGA) was chosen for the exchange-correlation functional. The effect of strong correlation was incorporated via the effective interaction parameter $U_{\rm eff}$ = $U - J$ through the rotationally invariant Dudarev’s approach \cite{Dudarev1998}. The Brilluion zone integration were carried out using $8 \times 8 \times 4$ and $12 \times 12 \times 6$ k-meshes to achieve the self-consistency and to obtain the densities of states (DOS), respectively. The PAW basis functions include 2$s$ and 2$p$ orbitals for O;  4$f$, 5$d$ and 6$s$ orbitals for Ho; and 3$d$ and 4$s$ orbitals for Co and Mn.

To examine the magnetic structure, we have designed various magnetic configurations which are depicted through Panel-(a) of Fig. \ref{bulk_bands_evo}. In these configurations various antiferromagnetic arrangements have been considered on Ho such as A-type (AFM1), G-type (AFM2) and C-type (AFM4) while keeping ferromagnetic interaction between Co and Mn whereas in one of the configurations (AFM3), the Ho-Ho interaction is maintained to be ferromagnetic while keeping G-type antiferromagnetic arrangement for Co and Mn. In Table -\ref{DOS}, we list the relative energies of the five magnetic configurations as a function of $U_{\rm eff}$ for two cases. For the first case the onsite correlation strength is varied uniformly on Co and Ho ($U_{\rm Mn}$ = 0) whereas for the second case a moderate value is fixed on Mn ($U_{\rm eff}$ = 2 eV) while varying $U_{\rm eff}$ uniformly on Ho and Co. A primary electronic structure analysis of spin polarized GGA DOS ($U_{\rm eff}$=0) (see inset in Fig. \ref{bulk_bands_evo}) suggests that the exchange splitting for Mn is nearly the same as the crystal field splitting ($\Delta_{\rm ex}\approx\Delta_{\rm cr}$) \cite{PARIDA2018}. As a result, there are no partially occupied states at the Fermi level and a gap exists even without considering the onsite correlation effect. On the other side, for Co, $\Delta_{\rm ex}>\Delta_{\rm cr}$ and the Fermi level is partially occupied with the down-spin channel of $t_{\rm 2g}$ manifold which gives rise to metallicity in the system. Therefore, finite correlation strength is required on Co to open up a gap whereas $U_{\rm eff}$ on Mn is hardly required. Furthermore, the robustness of the magnetic structure is also analysed for a moderate value of $U_{\rm eff}$ (= 2 eV) on Mn.

As inferred from Table-\ref{DOS}, under independent electron approximation (GGA) a FM ground state is preferred which is contrary to the experimentally reported Curie Weiss behavior where the negative value of $\theta_{\rm RE}$ infers an antiferromagnetic ground state. Often such discrepancy in magnetically active transition metal oxides arises due to lack of appropriate measure of strong correlation effect. In this regard as discussed earlier we have carried out GGA + U calculations that reveal AFM2 configuration, where Ho stabilize G-type ordering while Co and Mn possess ferromagnetic ordering, forms the ground state. Due to the $P$2$_{1}$/$n$ crystal symmetry, the configurations that allow the magnetic coupling between Co/Mn and Ho spins are FM and AFM3. However, as AFM2 constitutes the ground state, the Ho spins remain uncoupled to that of Co and Mn. Furthermore, we find that any kind of antiferromagnetic ordering among Co and Mn results in higher energy configuration. For example, the configuration AFM3, where nearest neighbor Co and Mn are antiferromagnetically oriented, is at least two orders higher in energy as compared to the ground state. This further validates the positive value of $\theta_{\rm TM}$ suggesting  FM interaction as obtained experimentally. 

The local magnetic moments listed in Table-\ref{DOS} along with the orbital projected DOS, shown in Fig. \ref{bulk_bands_evo}, describes the electronic and magnetic structure of HCMO. As can be clearly seen, for Mn, irrespective of $U_{\rm eff}$, in the majority spin channel the $t_{\rm 2g}$ states are occupied and the $e_{\rm g}$ states are empty whereas in the minority spin-down channel the $d$ states are completely unoccupied inferring a 4+ charge state with the 3\,$\mu_{\rm B}$ spin moment arising from $t_{\rm 2g}^{3\uparrow} e_{\rm g}^{0}$ electronic configuration. For Co and Ho, the $d$ and $f$ states are completely occupied in the majority spin channels while in the spin minority channel they are partly occupied highlighting the 2+ and 3+ charge states with $t_{\rm 2g}^{3\uparrow}t_{\rm 2g}^{2\downarrow} e_{\rm g}^{2\uparrow}$ and $f^{7\uparrow 3\downarrow}$ electronic configurations. The onsite repulsion due to strong electron correlation creates a lower and upper Hubbard bands (LHB and UHB) in the minority Co-$d$ spectrum, as shown in Panel-(b) of Fig. \ref{bulk_bands_evo}, to create a narrow band gap insulating system. The charge states inferred from the electronic structure analysis are in very good agreement with the experimentally observed charge states from the XPS measurements. The stabilization of the Co$^{2+}$ and Mn$^{4+}$ electronic configurations in the octahedral environment matches with the hypothesis proposed by Parida \textit{et al.} \cite{PARIDA2018} and demonstrate the case where the exchange field is comparable or greater than the crystal field split.

\section{Conclusions}
We synthesized high-quality polycrystalline samples of the double perovskite Ho$_{2}$CoMnO$_{6}$, which crystallizes in monoclinic $P$2$_{1}$/$n$ space group. X-ray photoelectron spectroscopy  unveil the exact charge state of cations in this material. The magnetic hysteresis curves $ M (H)$ at 5 K confirms a FM state owing to superexchange interaction between 3$d$ moments Co$^{2+}$-O-Mn$^{4+}$.  The specific heat confirms the presence of
ferromagnetic long-range ordering. The bifurcation of ZFC-FC curves at $T_{\rm c}$ = 76 K in low field \textit{H} = 100 Oe, indicates the presence of spin-freezing in this double perovskite material, which is also supported  by the frequency dependent ac susceptibility data. Also, the Vogel-Fulcher and critical slowing model fits point towards the stabilization of a re-entrant spin-glass ground state below $T_{\rm f}$ = 30 K as the spin flipping has a longer relaxation time compared to conventional spin-glass. The presence of competing interactions and anti-site disorder lead to such a spin-glass state in HCMO. In addition, magnetization results suggest the existence of a disordered state and short-range spin correlations above $T_{\rm c}$ in this material. The temperature dependent Raman spectra supports the magnetization results. Our Raman spectroscopy results suggest the significant contribution of spin-phonon coupling in the magnetically ordered state that arises from the phonon mediated modulation of the exchange integral in this material. The presence of spin-spin correlations is manifested by the deviation of the spectral line-shift from its anharmonic nature around 85 K.  The relatively large value of magnetocaloric entropy change,  $\Delta S_{\rm m} \simeq $ 13.5 (J/kg-K)  much below the ferromagnetic ordering temperature  suggests that HCMO  is a promising candidate  for magnetic  refrigerant applications. The electronic structure calculations corroborate the experimentally observed charge states and provide insights into the magnetic ordering in this double perovskite. The ground state magnetic ordering is found to be the one where the Ho spins form G-type antiferromagnetic arrangement while Co and Mn spin sublattices constitute ferromagnetic ordering. Furthermore, our calculations reveal that the present double perovskite is a narrow band gap insulator which is consistent with XPS results. Due to strong correlation effect in the spin minority channel, the Co-$t_{\rm 2g}$ manifold splits into lower and upper Hubbard subbands to create a  narrow gap. The spin majority channel has a natural band gap due to the crystal field split of the Mn-$d$ states.
Further studies are desired  to understand the complex interplay between emergent degrees of freedom leading to interesting physical phenomena in this 3$d$-4$f$ based double perovskite that may shed deep insights relevant in the context of novel material with competing magnetic order.

\section{Acknowledgment}
SK acknowledges support from IIT Madras.  PK acknowledges the funding by the Science and Engineering Research Board, and Department of Science and Technology, India through Research Grants. B.R.K.N. would like to acknowledge the funding from Department of Science and Technology, India, through grant No. CRG/2020/004330.


\bibliographystyle{apsrev4-1}
\bibliography{All_Citation}

\end{document}